\documentclass[12pt]{iopart}
\usepackage{graphicx}
\usepackage[russian]{babel}
\usepackage{hyperref}
\usepackage{epigraph}

\begin{document}
\selectlanguage{russian}

\title{Physics and Metaphysics}
\author{E. A. Solov'ev}

\address{Bogoliubov Laboratory of Theoretical Physics, Joint Institute for Nuclear Research, 141980 Dubna, Moscow region, Russia}

\ead{esolovev@theor.jinr.ru}

\begin{abstract}
Experiment plays a decisive role in physics. It is the single source
of our understanding of nature. But, during the last century the
main accent in theoretical physics  has moved toward metaphysics.
Some mathematicians/theoreticians try to develop new approaches,
ignoring experiments. Three examples of such treatment are discussed
here: the Gutzwiller approach, the classical description of the
tunneling phenomenon and the role of irregular motion in classical
mechanics. The new interpretation of quantum physics and classical
electrodynamics as theories of 'global information field' is
proposed. In conclusion the foundations of special relativity are
also discussed.
\end{abstract}


\setlength\epigraphwidth{8.truecm}

\epigraph{Красота - не прихоть полубога

А хищный глазомер простого столяра

(In English: Beauty is no demigod's whim,

It's the plain carpenter's fierce rule-of-eye.)}{O.Mandelshtam}

\section{Introduction}\label{intro}

Metaphysics was formulated by Aristotle (384 BC-322 BC). Its basic
statement is that true knowledge can be obtained only through logic.
Metaphysics was the only way for investigation of the Universe until
Galileo Galilei (1564-1642), who formulated the {\it scientific}
approach to the study of inorganic nature - {\it Physics}. First of
all, Galileo Galilei discovered the fundamental role of experiments
in inorganic nature - physical events are reproduced at any place
and any moment under the same external conditions i.e., they obey
the laws of nature which take the form of mathematical equations
introduced by Isaac Newton. Thus, experiment is the single source of
the understanding of nature. But, during the last century the main
accent in theoretical physics moved toward metaphysics. Some
mathematicians/theoreticians try to develop new approaches based
only on logic, ignoring the experimental background of physics. Of
course, physical theories use mathematics. But mathematics is a
logical scheme related more to our mentality than to nature. In
principle, mathematics does not produce new information; its aim is
the {\it identity} transformation, according to its axioms, of the
initial expression to a form more transparent for our consciousness,
e.g., by solution of a differential equation. Thus, mathematics
plays a subordinate role in physics. Investigations based only on
the abstract mathematical background lead, sometimes, to artifacts.
To illustrate this statement, three examples of such metaphysical
approaches are presented below.

\section{The Gutzwiller approach}

The first example is the Gutzwiller 'theory'. In this approach, the
contribution of the unstable periodic orbit to the trace of the
Green function is determined by the  formula \cite{Gutz}
\begin{equation}
g(E)\sim -\frac{iT(E)}{2\hbar}\sum_{n=1}^{\infty}\frac{\exp\{in[S(E)/\hbar-%
\lambda\pi/2]\}}{\sinh[nw(E)/2]}  \label{4.1}
\end{equation}
where $S(E)$, $w(E)$, $T(E)$ and $\lambda$ are the action, the
instability exponent, the period and the number of focal points
during one period, respectively. After the expansion of the
denominator, according to
$[\sinh(x)]^{-1}=2e^{-x}\sum_{k=0}^{\infty}e^{-2kx}$, and
summation of the geometric series over $n$ one can see that the
response function (\ref{4.1}) has poles at complex energies
$E_{ks}$ whenever
\begin{equation}
S(E_{ks})=\hbar\lambda\pi/2-i\hbar w(E_{ks})\left(k+\frac{1}{2}%
\right)+2s\pi\hbar,  \ \ k,s=0,1,2,... , \label{4.2}
\end{equation}
which are treated as resonances of the concerned system. The
Gutzwiller approach was widely applied to different few-body
systems - the scattering of electrons on the Coulomb potential in the
presence of external magnetic \cite{Win87,Du87} and electric fields
\cite{Win89,Gr91}, and the scattering on the two-Coulomb-center potential
\cite{Gr91} {\it etc}. However, this approach has been introduced
{\it ad hoc}. It is based on local characteristics in the vicinity
of the periodic orbits ignoring the asymptotic region which is
responsible for the physical boundary condition. Thus, in this
scheme it is impossible to distinguish whether the energy belongs to
the discrete spectrum or the continuum. Besides, in nonseparable
systems unstable periodic orbits with long periods lie everywhere
dense in phase space (renormalization-group) and the response
function (\ref{4.1}) has a pathological structure like the
Weierstrass function which is continuous everywhere but
differentiable nowhere. The singularities predicted by expression
(\ref{4.1}) have no physical meaning in the case of the discrete
spectrum as well, since the energies $E_{ks}$ are complex.

\section{Classical description of tunneling phenomenon}

Recently, in \cite{Bend} it has been argued that "quantum" effects
can be explained in terms of classical trajectories. In particular,
the tunneling effect for a potential
\begin{equation}
V(x)=x^2/2-g x^3 \label{a}
\end{equation}
has been discussed. In this case, the lowest state is a
quasi-stationary state whose lifetime $\tau$, defined by the
population of the bound state $P(t)=e^{-t/\tau}$, can be
approximated for small $g$ using WKB theory by the expression
\cite{Bend2}

\begin{equation}
\tau=\frac{1}{2}g\sqrt{\pi}\exp(2/15g^2) \label{b}
\end{equation}
In Fig.8 of the paper \cite{Bend} (which coincides with insertion in
Fig.1 of present paper\footnote{Fig.1 and Tab.1 have been prepared
by Alexander Gusev}), a complex classical trajectory with lifetime
(\ref{b}) at $g=2/\sqrt{125}$  is represented with the comment
\\[0.5cm]
"... initially, as the particle crosses the real axis to the right
of the middle turning point, its trajectory is concave leftward, but
as time passes, the trajectory becomes concave rightward. It is
clear that by the fifth orbit the right turning point has gained
control, and we can declare that the classical particle has now
'tunneled' out and escaped from the parabolic confining potential.
The time at which this classical changeover occurs is approximately
at $t=40$. This is in good agreement with the lifetime of the
quantum state in (8), whose numerical value is about 20."
\\[0.5cm]
\begin{figure}
\begin{center}
\includegraphics[width=0.6\textwidth]{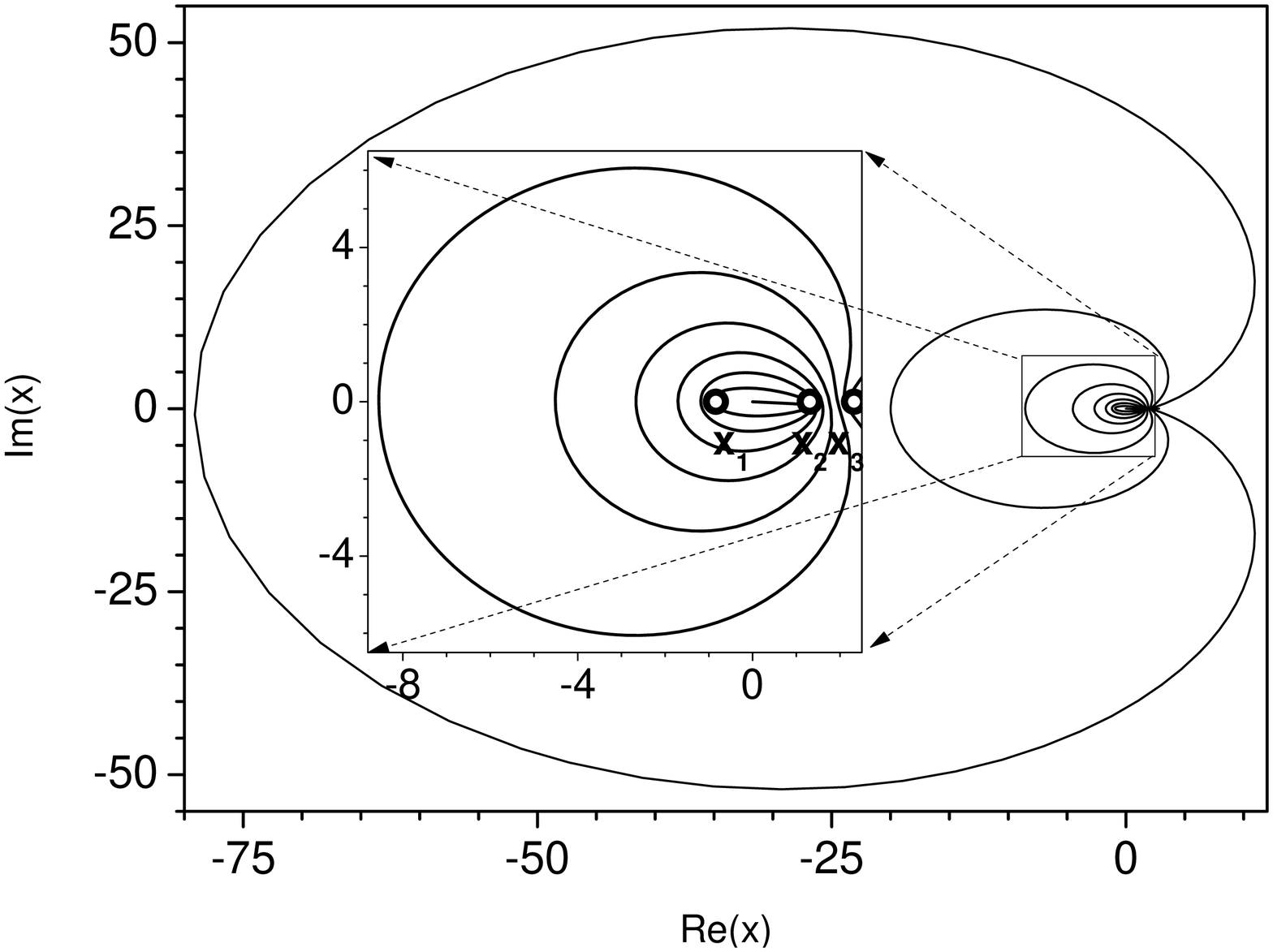}
\end{center}
\caption{The trajectory in the complex x-plane of a classical
particle with complex energy (\ref{b}) at $g=2/\sqrt{125}$. Open
circles in the insertion show the turning points $x_1$, $x_2$ and
$x_3$.}
\end{figure}

\noindent However, if the time increases further, one can see from
Fig.1 that the trajectory does not escape toward the right side but
continues to rotate mostly on the opposite side. On the other hand,
we can compare the quantum lifetime $\tau$ (\ref{b}), and the
classical time to reach the right turning point $t_c$ -
\begin{equation}
{\rm Re}\ x(t_c)={\rm Re}\ x_3
 \label{c}
\end{equation}
- proposed in \cite{Bend} as a classical analogue of $\tau$. It is
seen from Table 1 that there is nothing in common between these two
quantities in spite of  the "good agreement" declared in
\cite{Bend}. Thus, the effect discussed  is an artificial result
which has no relation to quantum theory.
\begin{table}[ht]
\caption{\label{tabone}The classical time $t_c$ (\ref{c}) and the
quantum lifetime $\tau$ (\ref{b}) versus coupling constant $g$. }

\begin{center}
\lineup
\begin{tabular}{*{5}{l}}
\br $g$ & 0.12522& 0.14311& 0.16099 & 0.17888 \cr \mr
$t_c$&\015009&\0\01385 & \0\0\0220 & \0\0\0\049 \cr
 $\tau$ & \0\0\0547& \0\0\0\085 & \0\0\0\024& \0\0\0\010\cr \br
\end{tabular}
\end{center}
\end{table}

\section{Chaos and irregular motion}

The next example is the study of irregular motion in classical
mechanics. The main point in this approach is the interpretation of
bifurcation as a source of chaos. Interest in the chaotic behaviour
is explained by an attempt to understand the origin of the arrow of
time provided by the Second Law of Thermodynamics, which says that
in an isolated system, entropy tends to increase in time. But there
is a fundamental contradiction between such evolution and classical
mechanics - the first is 100\% irreversible in time whereas
classical mechanics is 100\% reversible. Classical mechanics rather
deals with pseudo-chaotic evolution. This evolution is due to the
fundamental difference between rational and irrational numbers in
mathematics, which has no relation to nature. Generally speaking,
the axioms of mathematics always provide behaviour that is reversible in
time. We can obtain irreversible evolution only if we introduce a
stochastic concept like 'probability' which is beyond standard
mathematics.

\section{Quantum Physics and Classical Electrodynamics as Theory of
Global Information Field}

From the very beginning the interpretation of quantum theory has
gone in an erroneous direction. Already the denotation - quantum
{\it mechanics} - implies that it is an expansion of classical
mechanics toward the behavior of particles on the atomic scale.
Actually, quantum theory deals with a new object - a 'global
information field' $\Psi(\bf{r})$ (see
\cite{Sol09,Sol12})\footnote{Of course, this denotation is quite
symbolic since we have no direct experience and adequate vocabulary
in microworld. For instance, the word 'field' can be replaced by
word 'wave'.}. A global information field (GIF) plays the same
fundamental role as a material point ${\bf r}_i(t)$ (for $i$-th
material point) in classical mechanics and an electromagnetic field
$\{ {\bf E}({\bf r}), {\bf H}({\bf r})\}$ in classical
electrodynamics.  The radius-vector $\bf{r}$ in $\Psi(\bf{r})$ has
the same meaning as in the electromagnetic field $\{ {\bf E}({\bf
r}), {\bf H}({\bf r})\}$. At the moment of measurement, which plays
the role of 'interaction' between particle ${\bf r}_i(t)$ and GIF
$\Psi(\bf{r})$, the result of measurement is obtained by the
replacement ${\bf{r}}={\bf r}_i(t)$ in GIF like in the case of the
interaction of a charged particle with the electromagnetic field.
Thus, GIF is a substance additional to 'material points' but without
mass and energy. This statement is in contradiction with
wave-particle duality. In the present interpretation we have
material points characterized by ${\bf r}_i(t)$, and  the 'global
information field' $\Psi(\bf{r})$ which is subject to the wave-type
Schr\"odinger equation\footnote{For photons the role of the
Schr\"odinger equation is played by the Maxwell equations which
predict the propagation of the flux of photons. Here the
electromagnetic field $\{ {\bf E}({\bf r}), {\bf H}({\bf r})\}$ acts
as GIF.}. Thus, there is no duality because of two different
substances - the material point (e.g., electron) and GIF. The
existence of the global information field follows from the
Einstein-Podolsky-Rosen {\it experimentum crucis} which shows that
at the moment of measurement the propagation of quantum information
is at least $10^4$ times faster than the propagation of light
\cite{Sal,Sal2}, i.e. we have a new, non-material (without mass-type
characteristics) object - {\it the global information field}.

All experimental devices are designed on classical principles, and a
single measurement gives the fixed value of the {\it classical} (or
particle) quantity. After multiple repetition of measurement we
obtain the distribution of the classical quantity, according to GIF
$\Psi(\bf{r})$. The global information field itself is quite
deterministic. The probabilities arise at a measurement stage in the
transfer of the data from the quantum (GIF) to the classical
(particle) level. Though intuitively everyone understands what
measurement is, it cannot be formalized, i.e. be described in a
mathematical form. Somewhere measurement is considered as a
'collapse' of the wave function. However, such a word-play explains
nothing.

The concept GIF have something in common with the concept {\it
analyticity} in mathematics. Mathematics is complete in the complex
plane only. In the complex plane a very powerful tool appears -
analyticity. Namely, if an analytical function is known inside any
small region  \footnote{More accurate - if analytical function is
known on the countable set of points having accumulation point.},
then it is known everywhere. In paper \cite{Sol12} this global
property was named as a 'God phenomenon'. This mysterious property
has profound consequences in quantum physics. The solutions of the
Schr\"{o}dinger equation are analytical functions with respect to
the variables analytically entering in the Hamiltonian (it is the
standard case for the real quantum systems). Analyticity allows to
obtain asymptotic (approximate) results in terms of some singular
points in the complex plane which accumulate all necessary data on a
given process. For instance, in the case of integration of the fast
oscillating function by steepest descent method, the asymptotic
value of integral is expressed as a value of integrand at the saddle
point. In slow atomic collisions the cross-sections of inelastic
transitions are determined by branch-points of the adiabatic energy
surface at the complex internuclear distance \cite{Sol05}, and so
on.

Up to now ignorant interpretations of quantum physics exist and they
are intensively discussed in the literature. For instance, the
so-called 'hidden variable theory' (see, e.g., Wikipedia) which is
in total contradiction to the particle (electron or photon)
diffraction phenomenon in the experiment involving propagation of a
beam of non-interacting (or one by one) particles through a
crystalline grid (or two slits). In this experiment one particle
leaves one dot on the screen, and the distribution of dots has a
diffraction profile predicted by quantum theory (or classical
electrodynamics for photons). Can this distribution be explained in
terms of 'hidden variables theory'?

\section{Concluding remarks}

Here three examples have been presented. The same metaphysical aspect occurs in
some other approaches such as special relativity. At the basis of
relativistic mechanics lies a Minkowski space $\{x,y,z,ct\}$.
However this representation has well-known unresolvable paradoxes.
In the two-body problem we should use two sets of variables:
$\{x_1,y_1,z_1,ct_1\}$ for the first particle and
$\{x_2,y_2,z_2,ct_2\}$ for the second one (see,e.g.,\cite{Roh} p.188
where these variables are introduced without comments). What is the
meaning of $t_1$ and $t_2$? It is a scholastic question like "How
many devils can be located on a pin head?". It has no connection
with experiment or reality. Thus, the self-consistent relativistic
mechanics for few-body systems cannot be formulated. In this
context, the standard relativistic approach to the description of a
{\it single} particle looks more like a Kunstst{\"u}ck, than a
theory. In the Maxwell theory there is no such paradox, since the
electromagnetic field is the sole object. The next contradiction in
special relativity is the twin paradox. On the one hand special
relativity is kinematics since it connects coordinates and time in
two inertial frames of reference and mutual deceleration of time
measured in two different frames has no physical meaning. Then, the
time delay in the twin paradox can be connected with the stage when one of
the frames of reference is accelerated. This stage does not depend on how
long both frames of reference move uniformly. But the prediction of a time
delay from special relativity is proportional to this interval of
time. In the book \cite{Roh} on page 278 Fritz Rohrlich comments on
this situation:
\\[0.5cm]
"This result has been repeatedly stated in the form of an apparent
contradiction known as the {\it twin paradox} or {\it clock
paradox}. (Since the author has great difficulty constructing
paradoxes from clear mathematical facts, he will not attempt to do
so.)"
\\[0.5cm]
\noindent At the basis of both paradoxes lies the postulate that the
speed of light is the same for all inertial observers. In fact
this statement is beyond common sense; probably, it is out of our
mentality since we have no direct experience and adequate vocabulary
in this field. Here it is pertinent to quote the last paragraph from \cite{Sol12}
:
\\[0.5cm]
"A scientific approach is quite restricted because at its roots
ordinary language lies which is not certain and complete in
principle. For instance, it admit such paradoxes as "Can God create
a stone so heavy that he cannot lift it?". Another example that
demonstrates the incompleteness of vocabulary is the fact that
physics cannot be represented in Eskimo language, which has dozens
of words for different types of snow (snow that fell down yesterday,
snow on which a dog-sled passed, {\it etc.}); however, there is no
word 'snow', which is too abstract for them. But what is the level
of our mentality (language $\rightleftharpoons$ mentality)? However,
it is the sole tool for communication. In mathematics, the trace of
this incompleteness is G\"odel's theorem \cite{God}, which states
that there are true propositions about the natural world that cannot
be proved from the axioms."
\\[0.5cm]

\section*{ACKNOWLEDGMENTS}

I am grateful to Vladimir Belyaev, John Briggs, Tasko Grozdanov,
Vladimir Melezhik and Sergey Molodtsov for critical reading of the
manuscript.

\end{document}